\documentclass{article}

\usepackage{microtype}
\usepackage{graphicx}
\usepackage{subfigure}
\usepackage{booktabs} % for professional tables

\usepackage{hyperref}

\usepackage[accepted]{icml2025}
\setcitestyle{numbers,round,citesep={;},aysep={,},yysep={;}}

\usepackage{amsmath}
\usepackage{amssymb}
\usepackage{mathtools}
\usepackage{amsthm}

\usepackage[capitalize,noabbrev]{cleveref}

\theoremstyle{plain}

\theoremstyle{definition}

\theoremstyle{remark}

\usepackage[textsize=tiny]{todonotes}
\usepackage{graphicx}
\usepackage{amsmath,amssymb,amsfonts}
\usepackage{textcomp}
\usepackage{xcolor}
\usepackage{multirow}
\usepackage{multicol}
\usepackage{ifthen} 
\usepackage{caption}
\usepackage{subfigure}
\usepackage{comment}
\usepackage{setspace}
\usepackage{mathtools}
\usepackage{url}
\usepackage{listings}
\usepackage{comment}
\usepackage{makecell}

\usepackage{fontawesome5}
\usepackage{hyperref}

\newcommand{\MOD}{BIN}
\newcommand{\MODELNAMEFULL}{Bottleneck Iterative Network}

\renewcommand{\epsilon}{\varepsilon}
\DeclareMathAlphabet\mathbfcal{OMS}{cmsy}{b}{n} %

\newcommand{\ssh}[1]{\textcolor{blue}{[#1--- ss]}}

\icmltitlerunning{Submission and Formatting Instructions for ICML 2025}

\begin{document}

\twocolumn[
\icmltitle{Audio-Visual Speech Separation via Bottleneck Iterative Network}



\icmlsetsymbol{equal}{*}

\begin{icmlauthorlist}
\icmlauthor{Sidong Zhang}{umass}
\icmlauthor{Shiv Shankar}{umass}
\icmlauthor{Trang Nguyen}{umass}
\icmlauthor{Andrea Fanelli}{dolby}
\icmlauthor{Madalina Fiterau}{umass}
\end{icmlauthorlist}

\icmlaffiliation{umass}{Manning College of Information \& Computer Sciences, University of Massachusetts Amherst, Amherst, U.S.}
\icmlaffiliation{dolby}{Dolby Laboratories, San Francisco, U.S.}

\icmlcorrespondingauthor{Sidong Zhang}{sidongzhang@umass.edu}

\icmlkeywords{Machine Learning, ICML}

\vskip 0.3in
]



\printAffiliationsAndNotice{}  % leave blank if no need to mention equal contribution

\begin{abstract}
  Integration of information from non-auditory cues can significantly improve the performance of speech-separation models. Often such models use deep modality-specific networks to obtain unimodal features, 
  % which are statically combined to obtain high-level fused representations, 
  and risk being too costly or lightweight but lacking capacity. In this work, we present an iterative representation refinement approach called Bottleneck Iterative Network (\MOD{}), a technique that repeatedly progresses through a lightweight fusion block,
  % a technique that uses high-level fused representations to provide additional context to the lower-level network layers through backward connection, while controlling fusion expressiveness by using bottleneck fusion tokens. This helps improve the capacity 
  % expressiveness of the model,
  % while controlling nuisance representations in  fusion by bottleneck fusion tokens. 
  while bottlenecking fusion representations by fusion tokens.
  This helps improve the capacity of the model, while avoiding major increase in model size and balancing between the model performance and training cost. We test \MOD{} on challenging noisy audio-visual speech separation tasks, and show that our approach consistently outperforms state-of-the-art benchmark models with respect to SI-SDRi on NTCD-TIMIT and LRS3+WHAM! datasets, while simultaneously achieving a reduction of more than 50\% in training and GPU inference time across nearly all settings.   
\end{abstract}

\section{Introduction}
\vspace{-5pt}
The goal of speech separation is to separate a multi-speaker audio stream into multiple single-speaker audio streams. 
% This task is critical not only as a standalone application but also as an important first step in several downstream applications like speech enhancement and audio editing. 
% \trang{need citation} 
Pure audio-based methods for this speech separation task have long been studied in signal-processing \citep{hendriks2013dft,Luo_2019_AVConvtas}, but often face limitations when factors like reverberations, heavy overlap and background noise are common \citep{michelsanti2021overview}. Inspired by multimodal speech cognition in humans \citep{partan1999communication,mcgurk1976hearing}, contemporary research is pivoting towards a multimodal approach that integrates visual cues. The Audio-Visual Speech Separation (AVSS)  problem is about taking a single-channel audio stream of multiple speakers along with a corresponding video that captures all the speakers' faces, with the goal of identifying the utterance of each speaker \citep{Luo_2019_AVConvtas,lin2023av,AVLIT}. 

\textbf{Related Work.} 
AVconvTasnet \citep{Luo_2019_AVConvtas} extended the ConvTasNet  \citep{wu2019time} paradigm, an audio-only speech separation state-of-the-art model, to the audio-visual domain, setting one of the earlier baselines for the AVSS task. Since then, there have been various deep learning methods developed for the task \citep{lee2021looking,li2020deep}. \citet{gao2021visualvoice} used speaker images as external cues for separating audio using a large foundation model. RTFS-Net \cite{pegg2024rtfs} currently sets the frontier with respect to speech separation quality but comes with not only expensive training cost but also high inference time. This is a critical drawback because lots of important audio enhancement applications, such as real-time conference call enhancement, require fast speech separation. AVLIT \cite{AVLIT} and IIA-Net \cite{li2024iianet} \footnotemark are recent lightweight AVSS models. However, we find that there is a significant gap in output quality for noisy AVSS between these lightweight models and SOTA (state-of-the-art) RTFS-Net. 

  \footnotetext{IIA-Net has an additional configuration which achieves better output quality, but its inference time of 0.52s per 1 batch size sample and training time of ~300 hours on LRS3+WHAM! render it computationally too intensive and beyond the scope of this work.}

\textbf{Our Contribution.} We introduce a new AVSS model, Bottleneck Iterative Network (\MOD{}), that iteratively refines the audio and visual representations using their fused representation via a repetitive progression through the bottleneck fusion variables and the outputs of the two modalities from the same fusion block. Unlike\cite{iterative_Bralios_2023} which only iteratively refine unimodal latent features, 
\MOD{} refines both unimodal and fused representations, while bottlenecked by fusion variables after each iteration for robust performance \cite{nagrani2021attention}.
% \ssh{repetitive progression might make it seem that we have multiple layers}

Tested on two popular AVSS benchmarks, \MOD{} strikes a good balance between speech separation quality and computing resources, being on par with RTFS-Net's state-of-the-art performance (and improving on SI-SDR) while saving up to 74\% training time and 80\% GPU inference time. Our code is available on Github \href{https://github.com/Information-Fusion-Lab-Umass/BottleneckIterativeNetwork}{\faGithub}.

\label{sec:prelim}

\section{\MODELNAMEFULL{}}
\vspace{-5pt}

\begin{figure*}[th]
  \centering
  \includegraphics[width=0.85\linewidth]{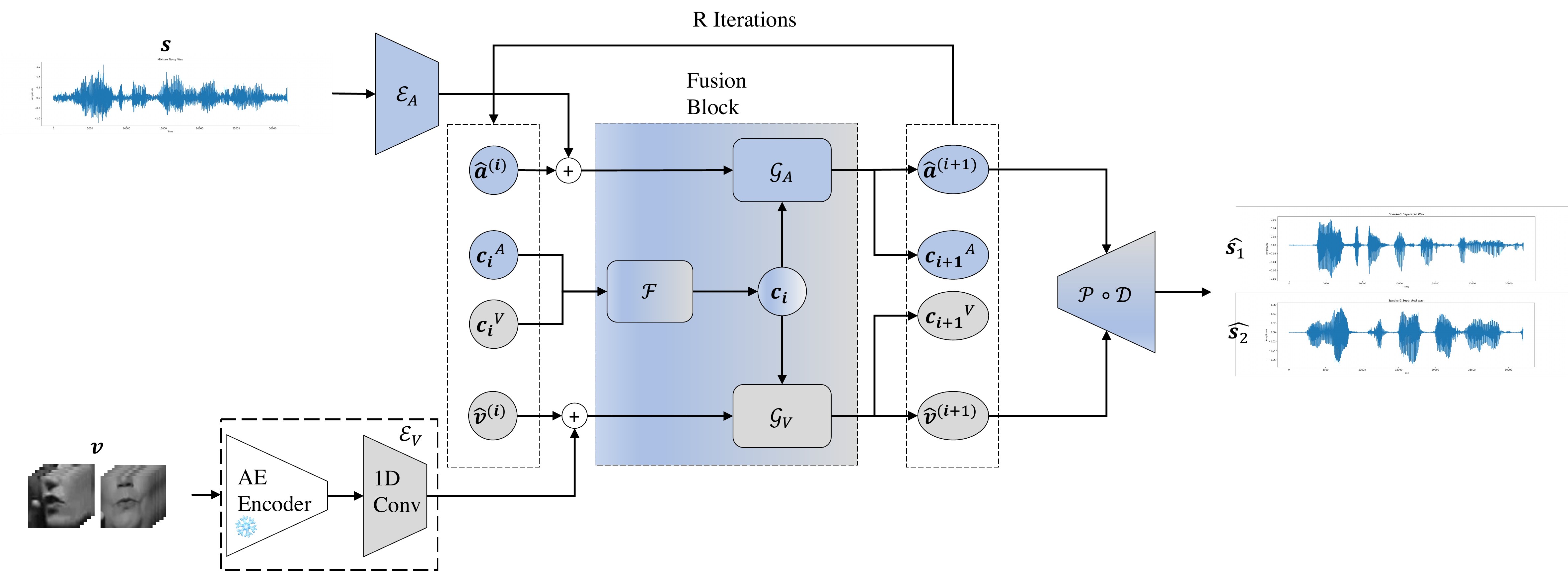}
  \caption{Bottleneck Iterative Network (\MOD{}) for Noisy AVSS model design.
  % \ssh{ I think the right part of the image with the gray block can be bigger and we can make the left part a bit less wide, or maybe just shrink the spectrogram thing. also that gray dotted line is a bit hard to see}
  }
  \label{fig:profusion-avss}
\end{figure*}

The noisy AVSS problem can be formulated as follows. There are two inputs: a noisy audio mixture record $\mathbf{s}$, which is composed of clean utterances of $M$ speakers $\mathbf{s}_{1...M}$ and background noise $\mathbf{n}$, and video input $\mathbf{v}$, which is a concatenation of video frames of lip regions $\mathbf{v}_1$ to $\mathbf{v}_M$ from the $M$ speakers. The expected outputs are a list of $M$ separated single-speaker audio streams $\mathbf{\hat{s}}_{1...M}$. 

\textbf{High-level justification}: We propose \MOD{} (illustrated in Figure \ref{fig:profusion-avss}) to leverage the downstream fused representation to refine the upstream audio and visual embeddings. At each iteration, the module takes the current unimodal embeddings and fusion tokens as input, and outputs updated fusion tokens along with refined unimodal latent representations. These refined representations, enriched with cross-modal information, are then added back to the original embeddings as residuals—effectively propagating downstream cross-modal context to the upstream unimodal features.

\textbf{Audio embedding model} $\mathcal{E}_A$: For an input audio mixture $\mathbf{s}$, a latent embedding space $\mathcal{E}_A(\mathbf{s})$ preprocesses the audio with an 1D convolution network $\mathcal{E}_A$. $\mathcal{E}_A(\mathbf{s})$ embeds $\mathbf{s}$ to a latent space of dimension $C_A \times F_A$, where $C_A$ expands the first dimension of $\mathbf{s}$ to a deeper hidden channel size, and $F_A$ compresses the time axis. $\mathcal{E}_A$ is only called once before the \MOD{} iteration for audio preprocessing.

% \textbf{Video embedding model} $\mathcal{E}_V$: The video embedding model is the encoder part of a 4-layer convolutional autoencoder pretrained for the video recover task. We freeze its parameters during training. A video sample $\mathbf{v}$ is of shape $N \times Fr \times C \times H \times W$, where $N$ is the number of speakers, $Fr$ is the total number of frames, $C$ is the RGB channel size, and $H \times W$ represents the height and width of each frame. The embedding space is of shape $C_V \times F_V$, where $F_V$ preserves a compacted sequential information along the time axis and $C_V$ is the latent dimension combining the number of the speakers, the RGB channel size, height and width for each frame. We further interpolate the video embedding so that $F_V = F_A$.

% \textbf{Video embedding model} $\mathcal{E}_V$: The video embedding model consists of the encoder of a 4-layer convolutional autoencoder pretrained for the video recover task, and a 1D convolution network. We freeze the autoencoder's parameters during training. A video sample $\mathbf{v}$ is of shape $N \times Fr \times C \times H \times W$, where $N$ is the number of speakers, $Fr$ is the number of frames, $C$ is the RGB channel size, and $H \times W$ represents the height and width of each frame. The embedding space is of shape $C_V \times F_V$, where $F_V$ preserves a compacted information along the time axis and $C_V$ is the latent dimension combining the number of the speakers, the RGB channel size, height and width for each frame. We further interpolate the video embedding so that $F_V = F_A$.

\textbf{Video embedding model} $\mathcal{E}_V$: The video embedding model consists of the encoder of a 4-layer convolutional autoencoder pretrained for the video recover task, and a 1D convolution network. We freeze the autoencoder's parameters during training. The embedding space is of shape $C_V \times F_V$, where $F_V$ compacts the temporal information and $C_V$ is the latent dimension combining the number of the speakers, the RGB channel size, height and width for each frame. We further interpolate the video embedding so that $F_V = F_A$.

% For our proposed method, $\mathcal{E}_V$ is a 4-layer convolutional autoencoder (AE). Each layer performs a 2D convolution with kernel size 2 and stride 2, followed by a LeakyReLU activation with slope 0.3. Following the AVLIT model \citep{AVLIT}, the video embedder is first pre-trained as an autoencoder on a video recovery task. This pre-trained autio encoder is then frozen and used to provide visual embeddings for training the AVSSS model. Since the number of video time fragmes and audio frames need not be aligned, we use a linear interpolation function to align the time axis of the output video, to obtain video embeddings corresponding to a fixed audio time.

%Same as audio, $\mathcal{G}_V$ is the preprocessing model for the video modality, and we follow the practice in \cite{AVLIT} to freeze its parameters during the Pro-Fusion training process.  

\textbf{Fusion Block}: There are four components in the Fusion Block. The Fusion Block is repeated for $R$ iterations, with shared parameters.% are shared among the blocks.

\textit{\textbf{Fusion variable}} $\mathbf{c}$: We utilize two learnable fusion variables $\mathbf{c}^A$ of shape $C_{AH} \times F_A$ and $\mathbf{c}^V$ of shape $C_{VH} \times F_V$, update them  with the partial output of the audio and video feature generator, and apply the variable fusion function on them, as explained later in Equations \ref{eqn:ga},\ref{eqn:gv},\ref{eqn:fusion}. 
% At every iteration, we update the two fusion variables by projecting partial output of the the audio and visual networks, a
% \ssh{ Someone should read the above paragraph and ensure it makes sense and is also correct wrt the code}

\textit{\textbf{Audio feature generator}} $\mathcal{G}_A$: At iteration $i$, $\mathcal{G}_A$ takes the generated audio feature from the previous iteration as the backward connection for refinement with a residual connection to $\mathcal{E}_A(\mathbf{s})$, a fusion variable $\mathbf{c}_{i-1}$ from last iteration for the bottleneck audio-visual fusion information, and outputs the new audio feature $\hat{\mathbf{a}}^{(i)}$ together with $\mathbf{c}_i^A$. Formally, 

\vspace{-0.5cm}
\begin{align}\label{eqn:ga}
\hat{\mathbf{a}}^{(i)} || \mathbf{c}_i^A = \mathcal{G}_A(\hat{\mathbf{a}}^{(i-1)} + \mathcal{E}_A(\mathbf{s}), \mathbf{c}_{i-1})
\end{align}

 We define $\hat{\mathbf{a}}^{(0)} = \mathbf{0}$.  At the first iteration, $\mathbf{c}_0$ is defined as the average of the learnable  $\mathbf{c}^A$ and $\mathbf{c}^V$.
 We used an Asynchronous Fully Recurrent Convolutional
Neural Network (A-FRCNN) \citep{afrcnn2021nips} as the audio feature generator on a convolved concatenation of $\hat{\mathbf{a}}^{(i-1)} + \mathcal{E}_A(\mathbf{s})$ and $ \mathbf{c}_{i-1}$.

% \trang{should specify what exactly happen in $\mathcal{G}$, is it just MLP?}

% \textit{\textbf{Video feature generator} }$\mathcal{G}_V$: We follow the same scheme as in $\mathcal{G}_A$, 
% % and produce $\hat{\mathbf{v}}^{(i)}$ together with $\mathbf{c}_i^V$. 
% Mathematically, at iteration i,

\textit{\textbf{Video feature generator} }$\mathcal{G}_V$: We follow the same scheme as $\mathcal{G}_A$ modeled by an A-FRCNN:
% and produce $\hat{\mathbf{v}}^{(i)}$ together with $\mathbf{c}_i^V$. 
% Mathematically, at iteration i,

 \vspace{-0.5cm}
\begin{align}\label{eqn:gv}
\hat{\mathbf{v}}^{(i)} || \mathbf{c}_i^V = \mathcal{G}_V(\hat{\mathbf{v}}^{(i-1)} + \mathcal{E}_V(\mathbf{v}), \mathbf{c}_{i-1})
\end{align}

% Similar to the audio-modality we use an A-FRCNN \citep{afrcnn2021nips} as the video feature generator.   

% \paragraph{Fusion and mapping function} $\mathcal{F}$, $\mathcal{W}_i$: We choose a simple fusion and mapping function to reduce the model complexity. For a given $\mathbf{c}_i^A$ and $\mathbf{c}_i^V$,

%  \begin{align}\label{eqn:fusion}
% \mathbf{c}_{i+1} = \mathcal{W}_i \circ \mathcal{F} (\mathbf{c}_i^A, \mathbf{c}_i^V) = \frac{1}{2}(\mathbf{c}_i^A + \mathbf{c}_i^V) 
% \end{align}

\textit{\textbf{Variable Fusion function}} $\mathcal{F}$: We choose a simple aggregation based fusion function \citep{khan2012color} to reduce model complexity. For a given $\mathbf{c}_i^A$ and $\mathbf{c}_i^V$,

 \vspace{-0.5cm}
\begin{align}\label{eqn:fusion}
\mathbf{c}_{i} = \mathcal{F} (\mathbf{c}_i^A, \mathbf{c}_i^V) = \frac{1}{2}(\mathbf{c}_i^A + \mathbf{c}_i^V) 
\end{align}

\textbf{Predictive function} $\mathcal{P}$ and \textbf{audio decoder} $\mathcal{D}$: A common practice to separate audio in AVSS is to predict the masks for the clean audio records \cite{wu2019time, AVLIT, li2024audiovisual}. Given the output $\hat{\mathbf{a}}^{(R)}$ and $\hat{\mathbf{v}}^{(R)}$ after $R$ iterations, a predictor $\mathcal{P}$ is a 1D convolution network mapping the latent output of A-FRCNNs back to the embedding space as the mask $\mathbf{m} = \mathcal{P}(\hat{\mathbf{a}}^{(R)}, \hat{\mathbf{v}}^{(R)}) $.
% takes both of them and generates the mask $\mathbf{m}$ of shape $N \times T$ for all the speakers:
% \begin{align}
% & \mathbf{m} = \mathcal{P}(\hat{\mathbf{a}}^{(R)}, \hat{\mathbf{v}}^{(R)}) \label{eqn:mask}
% \end{align}
Then the 1D transposed convolution network $\mathcal{D}$ retrieves the clean audio from the masked embedding audio:

\vspace{-0.5cm}
\begin{align}
    & \hat{\mathbf{s}_i} = \mathcal{D}(\mathcal{E}_A(\mathbf{s}) \odot \mathbf{m})[i-1], i \in \{1, 2, \cdots,  M\} 
\end{align}

where $\odot$ represents the element-wise product. 

% \ssh{reference to algo.. are we keeping algo or mergin the entire thing into a single model desc?}
% The major difference between the \MOD{} AVSS model and algorithm (\ref{algo: profusion})

% \sidong{currently I am removing all algorithm reference}

% \sidong{Merge the detailed implementation here}
%A major design choice in \MOD{} is that we restrict the cross-modality information exchange between audio and visual features to happen through only a narrow channel, extracting only task-relevant information. The bottleneck is important when certain modality inputs have a significant level of nuisance. Bottlenecking layers has been shown effective for regularization and has a compression and distillation effect \cite{srinivas2021bottleneck}. Previous research \citep{nagrani2021attention} has shown that bottlenecking cross-attention tokens in transformers leads to more robust performance multimodal action recognition tasks.

% \sidong{relation to avlit}

Following AVLIT \cite{AVLIT}, we use the lightweight A-FRCNN as feature generators instead of multiple recurrent networks in RTFS-Net \cite{pegg2024rtfs}. Unlike AVLIT which directly fuses audio and visual modality, we use the more computationally efficient bottlenecked fusion mechanism. We restrict the cross-modality information exchange between audio and visual features to happen through only a narrow channel, extracting only task-relevant information. The bottleneck is important when certain modality inputs have a significant level of nuisance. This bottleneck also enables our model to be trained and deployed within a manageable time budget despite having multiple iterations over the fusion blocks.
% If we set $R = 1$ and remove all the fusion tokens, our model has a similar structure to AVLIT \citep{AVLIT}. % One can argue that \MOD{} is a more general model structure to enhance the lightweighted AVLIT's performance.

% \trang{this phrasing reduces the importance of our method; how about "to allow more flexibility in utilizing the combined audio-visual representation and to enhance AVLIT performance"}

%, and we demonstrate that the concept of fusion with bottleneck is also applicable to the Pro-Fusion scheme in a model-agnostic manner.

% \ssh{We will need the figure of the neurl network being used for the seperation task. plus the model description similar to avlit or rtfn}

\section{Experiments}
\label{sec:expts}

% \ssh{shorten data desc and move the hyperparam details to appendix}
\subsection{Datasets}
\label{sec: avss-data}
\vspace{-5pt}
% We experiment on two following datasets:
%which to the best of our knowledge in the current state-of-the-art (SOTA) on noisy AVSS tasks.
%\footnote{RTFSNet \citep{pegg2023rtfs} has reported better results than AVLIT, however, they use non-noisy data which is less realistic and makes the results non-comparable}.
% \textbf{NTCD-TIMIT} is originally proposed for audio-visual automatic speech recognition task \citep{ntcd-timit}, where clean audio records from TCD-TIMIT dataset \citep{tcd-timit} are mixed with noise of six different types from the NOISEX-92 database\citep{NOISEX-92}. The Noise level is sampled uniformly randomly from a discrete list of -5db, 0db, 5db, 10db, 15db, and 20db, following the same practice in \citet{ntcd-timit} and \citet{AVLIT}. To create mixture audio records with noise, we use the same data splitting method in \citet{AVLIT}, where 3823 pairs of audio records sampled from 39 speakers are mixed as the training set, 784 pairs of audio records sampled from 8 speakers as the validation set, and 882 pairs of audio records sampled from 9 speakers as the testing set. Each audio record lasts 4 seconds at 16
% kHz sampling rate. For every audio mixture, a 4-second noise record is randomly sampled from the NOISEX-92 database to be added to the audio mixture. Since \citet{AVLIT} does not provide the assignment of every audio mixture record to a noise record, we match them randomly using our own script, thus the dataset we use in our experiment is not identical to theirs.

\textbf{NTCD-TIMIT:} Following \citet{AVLIT}, we mix clean audio records from TCD-TIMIT dataset \citep{tcd-timit} with noise sampled uniformly from the  NOISEX-91 database \citep{NOISEX-92} with varying loudness level of -5db, 0db, 5db, 10db, 15db, and 20db. Each record lasts 4 seconds at 16 kHz sampling rate. NTCD-TIMIT consists of 5 hours of training data, 1 hour of validation data and 1 hour of testing data. 

% \textbf{LRS3 +WHAM!} uses LRS3 \citep{lrs3} dataset to generate the audio mixture records of two speakers, and sample noise record from WHAM! \citep{wichern2019wham} dataset to create a noisy audio mixture. LRS3 dataset collects face tracks from over 400 hours of TED and TEDx videos, along with the corresponding subtitles and word alignment boundaries \cite{lrs3}. The WSJ0 Hipster Ambient Mixtures (WHAM!) dataset pairs each two-speaker mixture in the WSJ0-2M ix dataset \citep{hershey2016deep} with a unique noise background scene \cite{wichern2019wham}. In our experiment, we follow \citet{AVLIT} to generate 50,000 pairs of audio records from LRS3 as the training set, 5000 pairs of audio records as the validation set, and 3000 pairs of audio records as the testing set. Every LRS3 record is cut to 2-second long clip at 16
% kHz sampling rate before mixture. Since, \citet{AVLIT} does not clarify how they match LRS3 audio mixtures to WHAM! noise, so we split the noise audio records in WHAM! to 2-second clips, and assign them to the 58,000 audio mixture records without repeating, with noise levels sampled uniformly randomly from -6dB to 3dB. %Therefore, we also claim that the LRS3 +  WHAM! dataset in our experiment has the identical clean audio mixture, but very different noise distributions, compared with \citet{AVLIT}.
\textbf{LRS3 +WHAM!: } We follow \citet{AVLIT} to generate 50,000 pairs of audio records from LRS3 \cite{lrs3} as the training set, 5000 pairs of audio records as the validation set, and 3000 pairs of audio records as the testing set. Every LRS3 record is cut to 2-second long clip at 16 kHz sampling rate then mixed with noise sampled from WHAM! \cite{wichern2019wham} dataset.

\subsection{Evaluation}
\vspace{-5pt}
% We use Scale-invariant signal-to-distortion ratio improvement (SI-SDRi) metric as the primary evaluation metric to evaluate the output speech. SI-SDRi is computed as the SI-SDR difference between the mixture-separated audio pair and the mixture-clean audio pair, which reflects the speech quality improvement brought by the separation. For completeness, we also report two other metrics: a) Perceptual evaluation of speech quality (PESQ) is a common evaluation measure of speech quality \cite{pesq}  and b) Extended short-time objective intelligibility (ESTOI) \cite{estoi} measures the intelligibility of time-frequency weighted noisy speech with additional robustness to modulated noise \cite{Van_Kuyk_2018}. 

We use Scale-invariant signal-to-distortion ratio improvement (SI-SDRi) metric as the primary criterion to evaluate the output speech because it reflects the noise reduction and speech improvement brought by the separation. We also report two other metrics: a) Perceptual Evaluation of Speech Quality (PESQ) \cite{pesq}; b) Extended Short-time Objective Intelligibility (ESTOI) \cite{estoi}.
\vspace{-5pt}
\subsection{Performance Results}
\label{sec:performance}
\vspace{-5pt}

\begin{table*}[htb!]
\centering
\resizebox{0.9\textwidth}{!}{
\begin{tabular}{|l|l|ccc|ccc|}
\hline
Dataset & Model  &  \textbf{SI-SDRi} $\uparrow$ &  PESQ $\uparrow$ & ESTOI $\uparrow$ & \makecell{Training \\ time (h)}$\downarrow$ & \makecell{Peak GPU \\ Training \\ Memory (GB)} $\downarrow$ & \makecell{Peak GPU \\ Inference \\ Memory (MB)} $\downarrow$\\ 
\hline
\hline
\multirow{6}{*}{NTCD-TIMIT} 
& AVConvTasNet *     &  9.02  & 1.33  & 0.40 & -     & -  & -  \\
& IIA-Net **         &  6.04  & 1.33  & 0.38 & 2.92  & 1.03 & 148.82 \\
& AVLIT              &  8.59  & 1.39  & 0.44 & 3.88  & 2.98 & 238.15  \\
& RTFS-Net **        & 11.28  & 1.78  & 0.58 & 23.17 & 11.37 & 771.52 \\
\cline{2-8}
& \MOD/8 Iterations  & 10.68  & 1.51  & 0.50 & 5.45  & 3.88 & 269.44 \\
& \MOD/12 Iterations & 10.87  & 1.51  & 0.51 & 7.92  & 5.72 & 269.44\\
& \MOD/16 Iterations & \textbf{11.62} & 1.57 & 0.53 & 12.71 & 7.56 & 269.44  \\
\hline
\hline
\multirow{6}{*}{LRS3+WHAM!} 
& AVConvTasNet *     & 6.21  & 1.29  & 0.60 & -     & -  & -  \\
& IIA-Net **         & 8.93  & 1.36  & 0.55 & 16.74 & 0.58 & 140.57 \\
& AVLIT              & 11.62 & 1.53  & 0.65 & 28.09 & 1.54 & 159.48\\
& RTFS-Net **        & 12.14 & 1.74  & 0.70 & 193.45 & 5.66 & 414.58 \\
\cline{2-8}
& \MOD/8 Iterations  & 11.82 & 1.55  & 0.66 & 34.55 & 2.01 & 187.30\\
& \MOD/12 Iterations & \textbf{12.25} & 1.59 & 0.68 & 50.58 & 2.92 & 187.30 \\
& \MOD/16 Iterations & 10.84 & 1.49  & 0.53 & 81.91 & 3.84 & 187.30\\
\hline
\end{tabular}
}
\caption[test]{Performance on NTCD-TIMIT and LRS3+WHAM!. 
* results are as reported in earlier literature. \\
** these models were originally designed for speech extraction task and then adapted by us to speech separation task.}
\label{tab:avss}
\end{table*}

Table \ref{tab:avss} reports the evaluation results by comparing the audio separation qualities on the testing split of the two datasets using the benchmark models and proposed \MOD{} model, as well as their training costs on NVIDIA A100 GPU. We also investigate complexity of the models at inference stage in Table \ref{tab:complexity} by running on NVIDIA 2080 Ti GPU on a sample input of batch size 1 from LRS3+WHAM!. 

% \vspace*{-\baselineskip}
\begin{table}[htb!]
\resizebox{0.45\textwidth}{!}{
\begin{tabular}{|l|l|l|l|}
\hline
Model   & MACs (G) & \#Params (M) &  \makecell{GPU \\ Inference \\ Time (s)} \\
\hline
% VisualVoice     & 19.70    & 77.75       & 0.20               \\

IIA-Net & 9.93 & 3.52 & 0.03 \\
AVLIT         & 36.76    & 5.72         & 0.02 \\
RTFS-Net & 67.38 & 1.07 & 0.16 \\

% ProFusion-2-256 &  11.990 & 6.048 & 0.0073          \\
% ProFusion-8-128 & 40.840  & 5.88       & 0.026          \\
 \MOD/8 iterations & 42.95 & 6.05 & 0.03 \\
% ProFusion-8-512 & 47.155 & 6.377 & 0.0291 \\
% ProFusion-12-128 & 60.423 & 5.883 & 0.0401 \\
\MOD/12 iterations & 63.58 & 6.05 & 0.04 \\
% ProFusion-12-512 & 69.896 & 6.377 & 0.0435 \\
\MOD/16 iterations & 84.22 & 6.05 & 0.06 \\
\hline
\end{tabular}
}
\caption{More computational complexity on a sample of batch size 1 from LRS3+WHAM!.
% \ssh{maybe mention gpu,os,cuda no here or in appendix}
% \tablefootnote{\label{note1} These results will be added in the final version}
%\textsuperscript{\getrefnumber{note1}}
}
% We report model complexity and inference time. ProFusion is marginally more intensive than AVLIT, but its parametric complexity and inference times are also better than baseline models like VisualVoice
\label{tab:complexity}
\end{table}
% \vspace*{-\baselineskip}

% \sidong{We cite the performance of the models marked by * from \citet{AVLIT}, since we fail to find any replication instructions from  \citet{AVLIT} on adapting them to this noisy two-speaker AVSS task. Although the noise level in our datasets is different from \citet{AVLIT}, the replicated result of AVLIT is lower than what is reported in the original paper, indicating that the noise level in \citet{AVLIT} is possibly weaker.  }

% \ssh{check if this footnote is correct
%\footnotetext{The replicated result of AVLIT in our experiment is comparable but different from what is reported in the original paper, because the  noise distribution is different from that in \citet{AVLIT}. This is because there are no standard settings of the noise. We follow the same noise dB range as described in \citet{AVLIT} but noise sample using our own scripts, yielding different noise sample and noise dB level for every audio mixture sample.}
% }

% \ssh{some more hyperparameters are here, can skip that}
% \sidong{hyperparameters may still change}

  We observe that for both datasets, \MOD{} with different number of iterations brings performance improvement to all the three evaluation metrics compared with the lightweight AVLIT and IIA-Net performance from our replication study. Our most lightweight model, \MOD/8 iterations, has computational complexity comparable to AVLIT, while yielding slight performance improvement on LRS3+WHAM! and significant improvement on NTCD-TIMIT.

  Compared to RTFS-Net, our \MOD/16 iterations and \MOD/12 iterations have better performance with respect to SI-SDRi on NTCD-TIMIT and LRS3+WHAM!, while using only 55\% and 26\% of RTFS-Net training time, 66\% and 52\% of RTFS-Net training peak GPU memory, and 35\% and 45\% of RTFS-Net inference peak GPU memory.  \MOD{} also cuts the inference time by 80\% and 75\% on the two datasets. 

 % \sidong{Compare ours with rtfs here}
 % It is also worth noting that the motivation example in section \ref{sec:motiv} is a data sample from LRS3+WHAM!, where we initially observe that AVLIT fails to separate the beginning of the utterance and still outputs distorted separated audio records.  \MOD{}, on the other hand, can correctly separate the audio mixture to "WE MUST ADOPT THAT WAY OF" and "THEY'RE LIKE THE BASEBALL SCOUTS 20 YEARS".
% \vspace{-5pt}

\begin{figure}[htb!]
 \centering
\includegraphics[width=0.9\linewidth]{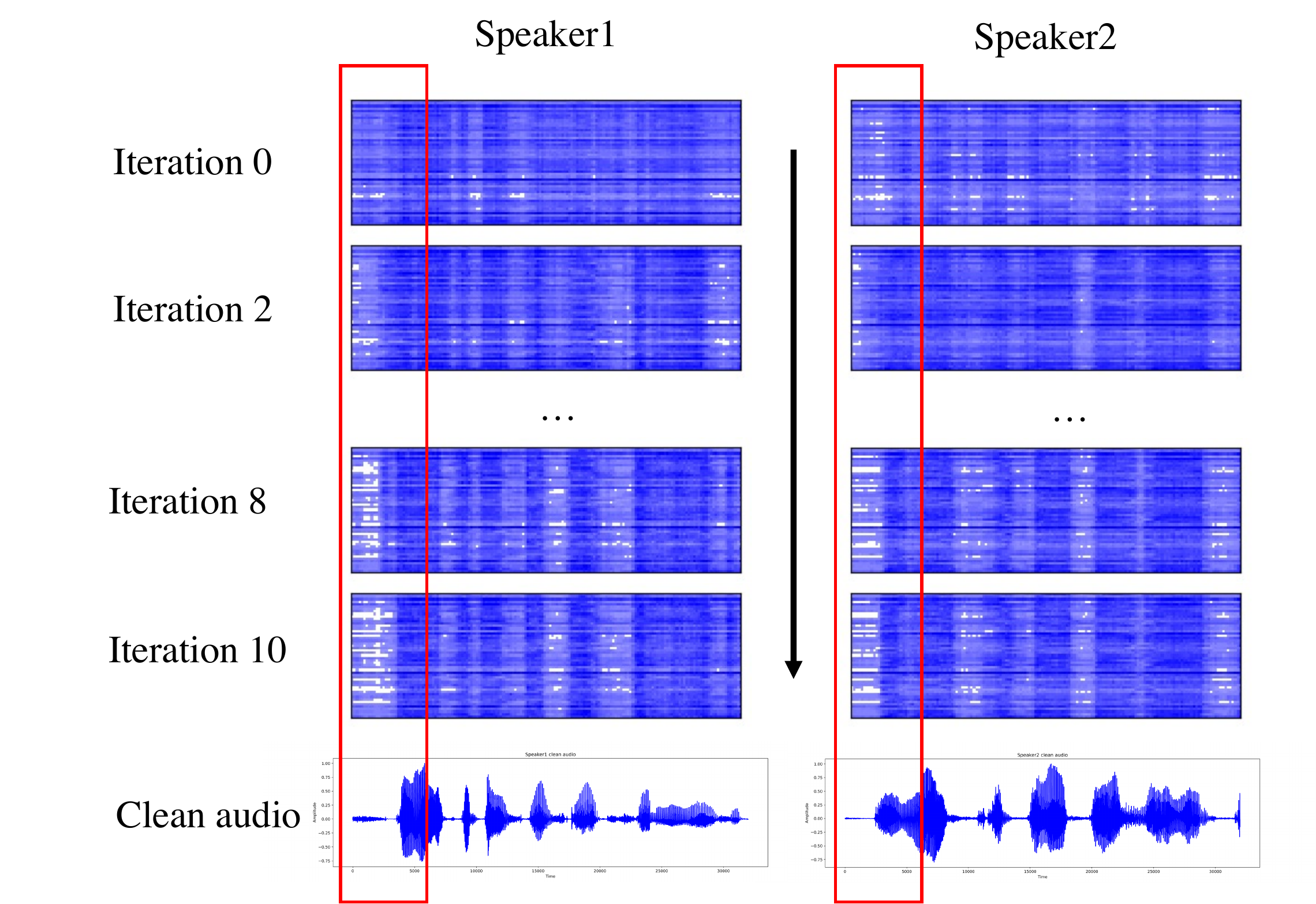}
 \caption{The progression of latent masks across fusion iterations of the final trained \MOD{} model on one LRS3+WHAM! test sample. The masks in the later iterations show patterns more aligned with the actual clean audio.} 
 \label{fig:iter-mask}
\end{figure}
% \vspace*{-\baselineskip}

% \vspace*{-\baselineskip}
\begin{figure}[htb!]
 \centering
\includegraphics[width=\linewidth]{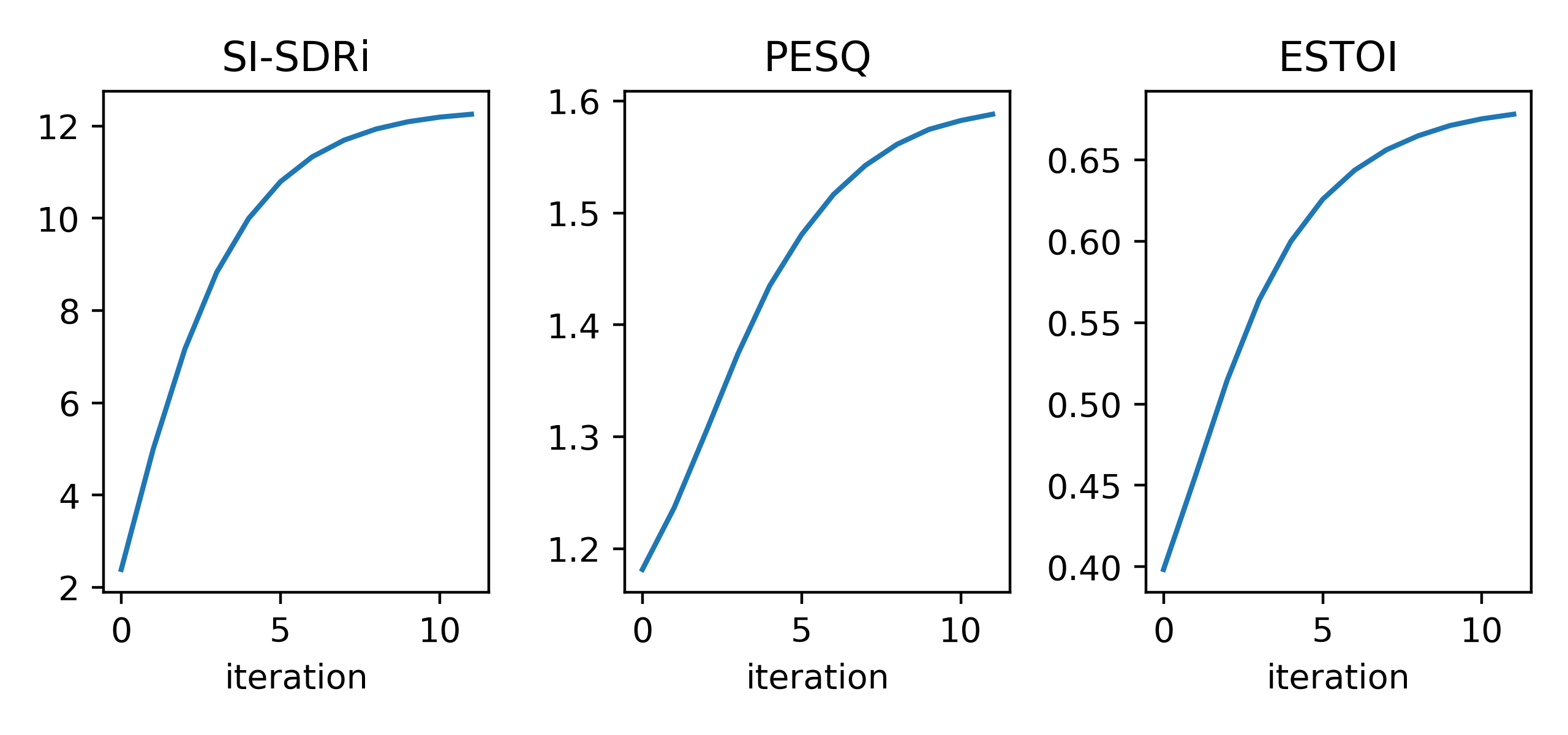}
 \caption{The progression of the audio separation quality through \MOD{} iterations in the trained \MOD{} model on LRS3+WHAM! testing split. %The iterations are not epochs of training, but the unrolling iterations $R$ of the model during test time.
 } 
 \label{fig:iter-performance}
\end{figure}
% \vspace*{-\baselineskip}

% \vspace{-5pt}
\subsection{Analysis of Iterative Design}
% \vspace{-5pt}
%of the Progressive Iteration}
% \begin{figure}[h]
%   \centering
%   \includegraphics[width=\linewidth]{figACMMM/iter_mask.pdf}
%   \caption{The progression of the hidden masks through Pro-Fusion iterations for the 2 separated audio records processed by the trained Pro-Fusion model. The mask lies in a latent space, but it shares a similar pattern of peaks and troughs to the clean audio record. The peaks and troughs are evolving along with the iteration, indicating the model is gradually accumulating information for the separated audio.}
%   % \ssh{can we also add the ideal mask corresponding to ground truth, here instead of the clean audio waveform}
%   % \sidong{idk how to do this}
%   \label{fig:iter-mask}
% \end{figure}

% \ssh{maybe we need a line here from rebuttal about it not being training evolution}
% We show the effect of running the $R$ iterations (i.e. the repeated application) of the fusion block, by visualizing the latent masks of progressive \MOD/12 iterations for LRS3+WHAM!, after finishing training the model on one test sample input. The iterations are not epochs of training, but are unrolling iterations $R$ of the model at test time. The mask lies in the latent space, but it shares matching peaks and troughs with the clean audio track. The peaks and troughs change as the number of iteration increases, indicating that the model is progressively optimizing the mask through the iterations. 

We next study how \MOD{} refines the task across multiple fusion iterations. Using the \MOD/12 iterations model trained for LRS3+WHAM!, we compute the audio separation masks $\mathbf{m}^{(i)}$ at each fusion iteration by applying the predictive function $\mathcal{P}$ to the intermediate audio and video features generated at each fusion iteration $\hat{\mathbf{a}}^{(i)}$ and $\hat{\mathbf{v}}^{(i)}$. In Figure \ref{fig:iter-mask}, we visualize the mask $\mathbf{m}$ as generated after each iteration along with the clean signal for a sample mixture. It shows the peaks and troughs become sharper in estimated masks of later iterations, matching the patterns in the clean speaker-specific audio. This qualitatively indicates that the model is progressively refining the mask through the fusion iterations. 

Quantitatively, we demonstrate the progression of audio separation  across fusion iterations on LRS3+WHAM! on the trained model at the inference stage in Figure \ref{fig:iter-performance}. We observe an initial sharp and then steady increase in output quality throughout the twelve iterations in \MOD/12 iterations.  
% \ssh{ this described a figure which may not keep in the draft, move to supplement}
% For the final trained \MOD{} model, we also present the progression of the audio separation quality through \MOD{} iterations on the testing split of LRS3+WHAM! (figure \ref{fig:iter-performance}). The curves clearly indicate the refinement of the audio separation through the progressive iterations. We also see that the performance saturates as we increase the number of iterations, with an initial sharp linear improvement in separation quality for low iterations. 

% \ssh{what expts from the rebuttal can come here. we need MACs but what else ? Feels COMPLETED}

\subsection{Analysis of Fusion Design}
% \vspace{-5pt}
% \begin{table}
% \begin{tabular}{c|l|lll}
% \hline
%   &  Model   & SI-SDRi & PESQ & ESTOI\\
% \hline
%  % \multirow{ 4}{*}{Fusion Position} & AVLIT EarlyFusion & 11.622 &1.527 & 0.653 \\
%  % & AVLIT LateFusion & 9.946 & 1.405 & 0.595\\
%  % %& AVLIT Middle {2} 1.30 0.56 9.15
%  % & EarlyFusion-12 & 11.983 & 1.573 & 0.668 \\
%  % & LateFusion-12 & 8.812 & 1.375 & 0.568 \\
%   \multirow{ 2}{*}{Fusion Pos.} & 
%  %& AVLIT Middle {2} 1.30 0.56 9.15
%   EarlyFusion & 11.983 & 1.573 & 0.668 \\
%  & LateFusion & 8.812 & 1.375 & 0.568 \\
% \hline
% \multirow{ 3}{*}{Ablation} & \MOD -No $\mathbf{c}^{A}$ & 11.534 & 1.535 & 0.652 \\
% & \MOD -No $\mathbf{c}^{V}$ & 11.556 & 1.528 & 0.649 \\
% % & ProFusion-No $\mathbf{c}^R$ & 8.812 & 1.375 & 0.568\\
% & \MOD -No $\mathbf{c}^R$ & 9.641 & 1.426 & 0.594 \\
% \hline
% \end{tabular}
% \caption{Ablation }
% \label{tab:ablation}
% \end{table}
% \trang{need to add params and full VAIS performance to Table III}

% \vspace*{-\baselineskip}
\begin{table}
\resizebox{0.4\textwidth}{!}{
\begin{tabular}{|l|c|cc|}
\hline
  Model   & SI-SDRi & PESQ & ESTOI\\
\hline
 % \multirow{ 4}{*}{Fusion Position} & AVLIT EarlyFusion & 11.622 &1.527 & 0.653 \\
 % & AVLIT LateFusion & 9.946 & 1.405 & 0.595\\
 % %& AVLIT Middle {2} 1.30 0.56 9.15
 % & EarlyFusion-12 & 11.983 & 1.573 & 0.668 \\
 % & LateFusion-12 & 8.812 & 1.375 & 0.568 \\

 %& AVLIT Middle {2} 1.30 0.56 9.15
\MOD -No Bottleneck & 11.99 & 1.57 & 0.67 \\
%  LateFusion & 8.812 & 1.375 & 0.568 \\
\MOD -No $\mathbf{c}$ & 9.64 & 1.43 & 0.60 \\
\MOD -No $\mathbf{c}^{A}$ & 11.53 & 1.53 & 0.65 \\
\MOD -No $\mathbf{c}^{V}$ & 11.55 & 1.52 & 0.65 \\
% & ProFusion-No $\mathbf{c}^R$ & 8.812 & 1.375 & 0.568\\
\MOD{} (Full) & \textbf{12.25}     &   1.59     &    0.68 \\
\hline
\end{tabular}
}

\caption{Ablation of \MOD/12 iterations on LRS3+WHAM!.}
\label{tab:ablation}
\end{table}
% \vspace*{-\baselineskip}

We investigate various modifications of the reported 
\MOD/12 iterations with respect to the fusion mechanism on LRS3+WHAM! and report results in Table \ref{tab:ablation}. The first row report the model performance without bottleneck fusion tokens. We use one single generator $\mathcal{G}$ on concatenated audio visual features, instead of two distinct feature generator functions $\mathcal{G}_A$ and $\mathcal{G}_V$. We can observe a performance decrease compared to \MOD, showing that using fusion tokens to restrict cross-modal information exchange is helpful. %However we also note that even the simplified model structure outperforms other lightweight models on noisy AVSS.

We further investigate the impact of multimodal information used to refine unimodal representations. \MOD -No $\mathbf{c}$ removes the variable fusion function $\mathcal{F}$ and $\mathbf{c}_i$, so the two tokens can only 
iterate through their own modality pipelines, which is similar to a late fusion design. Without cross-modal information exchange, the performance degrades significantly. \MOD -No $\mathbf{c}^A$ removes the audio fusion token and only allows the audio modality to receive video fusion tokens. Similarly, only the video modality can access the audio fusion tokens in \MOD -No $\mathbf{c}^V$. Due to the missing information from the other modality, performance is lower in both ablations.

\vspace{-5pt}
\section{Conclusion}
\vspace{-5pt}
Our paper presents a new approach for noisy AVSS that incorporates backward connections to reduce compute complexity, while retaining near SOTA performance. 
% In our experiments on two benchmark datasets designed for the noisy AVSS task, we show that \MOD{} can match and even slightly improve SOTA models performance on AVSS while cutting the GPU training time and GPU inference time significantly.
Our experiments on two noisy AVSS benchmark datasets show that \MOD{} can match and even slightly surpass the performance of SOTA models in AVSS while significantly reducing both GPU training and inference times. Future work will refine the fusion block design for better flexibility in complexity–performance trade-offs on larger scale AVSS tasks.

%, 
%Improvements in early identification of conditions such as Alzheimer's,  using information from health records, can add significantly to the quality of life for patients. A future potential direction of our work is to assess its applicability in healthcare domain. 

%\textbf{Limitations}: Our experiments do not give a clear answer to the question of how our approach interacts with other model agnostic methods. As \MOD{} instantiates a version of message passing on the graph in Figure \ref{fig:motiv}, if a task has a similar dependency structure, then our method can be expected to yield good improvements. That said, no model design is suitable for all problems, and our results might not generalize to other datasets.

% \begin{acks}
% \end{acks}

\begin{comment}
\section*{Acknowledgment}

The preferred spelling of the word ``acknowledgment'' in America is without 
an ``e'' after the ``g''. Avoid the stilted expression ``one of us (R. B. 
G.) thanks $\ldots$''. Instead, try ``R. B. G. thanks$\ldots$''. Put sponsor 
acknowledgments in the unnumbered footnote on the first page.
\end{comment}
%\pagebreak

%\section*{References}
\newpage
\bibliography{mybib}
\bibliographystyle{icml2025}
\end{document}